\documentclass[pdftex,twocolumn,epjc3]{svjour3}   
\RequirePackage[T1]{fontenc}

\smartqed  

\RequirePackage{graphicx}
\RequirePackage{mathptmx}      % use Times fonts if available on your TeX system
\RequirePackage{flushend}
\RequirePackage[numbers,sort&compress]{natbib}
\RequirePackage[colorlinks,citecolor=blue,urlcolor=blue,linkcolor=blue]{hyperref}
\RequirePackage{amsmath}
\RequirePackage{multirow}
\RequirePackage{xspace}
\RequirePackage{xcolor}
\RequirePackage{cancel}
\RequirePackage[caption=false]{subfig}
\RequirePackage{subfig}
\RequirePackage{blindtext, subfig}
\RequirePackage{amsmath}
\usepackage{amssymb}
\usepackage{dblfloatfix}% http://ctan.org/pkg/dblfloatfix
\usepackage{lipsum}% http://ctan.org/pkg/lipsum
\usepackage{ulem}
\RequirePackage{fix-cm}

\RequirePackage{comment}
\RequirePackage{hyperref}
\RequirePackage{slashed}
\RequirePackage[caption=false]{subfig}
\RequirePackage{tabu}
\usepackage{hyperref}
\usepackage{multicol}
\usepackage[super]{nth}
\usepackage{booktabs}
\usepackage{dutchcal} % \mathcal: also small letters. \mathbcal = bold ones.
\def\MGvATNLO{{\tt {\sc MadGraph5}\_aMC@NLO}}
\begin{document}

\title{Exploring dark $Z_d$-boson in future Large Hadron-electron collider}
 
\author{Ashok Goyal\thanksref{e1,addr1}
        \and
        Mukesh Kumar\thanksref{e2,addr2} %etc.
        \and
        Satendra Kumar\thanksref{e3,addr3} %etc.
        \and
        Rafiqul Rahaman\thanksref{e4,addr4} %etc.
}

\thankstext{e1}{e-mail: agoyal45@yahoo.com}
\thankstext{e2}{e-mail: mukesh.kumar@cern.ch}
\thankstext{e3}{e-mail: satendrak@iiitu.ac.in}
\thankstext{e4}{e-mail: rafiqulrahaman@hri.res.in}


\institute{Department of Physics, University of Delhi, Delhi, India. \label{addr1}
           \and
           School of Physics and Institute for Collider Particle Physics, University of the Witwatersrand, Johannesburg, Wits 2050, South Africa. \label{addr2}
           \and
           School of Basic Sciences, Indian Institute of Information Technology Una, Una, Himachal Pradesh  177 209, India.\label{addr3}
           \and
           Regional Centre for Accelerator-based Particle Physics, Harish-Chandra Research Institute, A CI of Homi Bhabha National Institute, Chhatnag Road, Jhunsi, Prayagraj 211 019, India.\label{addr4}
}

\date{Received: date / Accepted: date}
\maketitle

\begin{abstract}
The interaction between the dark $U(1)_d$ sector with the visible Standard Model (SM) sector takes place through the kinetic mixing between the dark photon $U(1)_d$ field $Z_d^\mu$ and the SM $U(1)_Y$ gauge field $B_\mu$. After the electroweak and $U(1)_d$ symmetry breaking, the dark photon $Z_d^\mu$ acquires a mass and mixes with the SM neutral vector boson $Z_\mu$. This mixing leads to parity-violating coupling between the $Z_d^\mu$ and SM. The coupling between the dark photon and SM can be explored in low energy phenomenology as well as in collider experiments. The Lorentz structure of dark photon interaction with SM fermions is explored in the proposed high energy future Large Hadron-electron collider, which would provide efficient energy and a clean environment using cross-section and asymmetries associated with polarisation observable of the dark photon in leptons decay. A $\chi^2$-analysis is performed to compare the strength of various variables for both the charge- and neutral-current processes. Based on this analysis, $90\%$ confidence level (C.L.) contours in the $\epsilon$-$m_{Z_d}$ and $\epsilon$-$g_V$ plane are obtained to put limits on the $Z_d^\mu$ mass up to $100$~GeV, coupling strength $\epsilon$ and on the Lorentz structure of dark photon coupling with the SM fermions ($g_V$) at $\sqrt{s} \approx 1.3$~TeV.
\end{abstract}

\section{Introduction}
\label{intro}

The existence of dark matter (DM) in the universe has been established by several cosmological and astrophysical observations. Dark matter contributes roughly $75\%$ of the entire matter existing in the universe. The nature of the DM has, however, remained undetermined so far. The negative results of the direct searches~\cite{Arcadi:2017kky} have reached a sensitivity level where they are in tension with the assumption that the DM has at least some common charge under the SM gauge interactions. This has led to consider a scenario in which there is a separate ``dark sector'' consisting of one or more particles that are not charged under SM gauge group. The dark and visible SM sectors interact through a portal - as it is commonly called. The portal consists of operators of different forms depending upon the spin of the mediator. One can have a scalar portal with Higgs boson, a pseudo scalar axion-like particle mediator, a fermionic heavy (sterile) neutrino mediator, or a vector mediator. These have been extensively studied, see for example Dark Sector 2016 Workshop Community Report~\cite{Alexander:2016aln} and Refs.~\cite{Batell:2017cmf,Blennow:2019fhy,Fabbrichesi:2020wbt}. In the case when the only interaction available to the dark-sector particles with the visible sector is gravitational, the mediator is a spin-2 graviton as in Kaluza-Klien (KK) warped extra-dimension models~\cite{Randall:1999ee, Kraml:2017atm, Lee:2013bua, Lee:2014caa, Rueter:2017nbk, Goyal:2019vsw}.

In the vector-portal, the interaction between the dark and visible sectors takes place through the kinetic mixing between the dark $U(1)_d$ and SM $U(1)_Y$ gauge fields $Z_d^\mu$ which is called ``dark photon" and $B^\mu$, respectively. The kinetic mixing is given by~\cite{Davoudiasl:2012ag} 
\begin{equation}
 {\mathcal L}_{\rm guage}
 = -\frac{1}{4} B_{\mu\nu} B^{\mu\nu} -\frac{1}{4} {Z_d}_{\mu\nu} Z^{\mu\nu}_d+\frac{1}{2} \frac{\epsilon}{\cos \theta_W} B_{\mu\nu} Z^{\mu\nu}_d
 \label{lgauge}
\end{equation}
with $V_{\mu\nu}=\partial_\mu V_\nu - \partial_\nu V_\mu,~V=Z_d/B$.
A redefinition of fields leads to an effective parity conserving induced coupling between the dark photon $Z_d$ and the electromagnetic current as
\begin{equation}
{\mathcal L}_{J_\mu} = - \epsilon\, e\, J_\mu^{em} Z_d^\mu. \label{ljmu}
\end{equation} 
The parameter $\epsilon$ is assumed to be small $\epsilon < {\cal O} (10^{-3}) - {\cal O} (10^{-1})$ but theoretically can be of order 1 and is not required to be small. It can be calculated in some beyond the SM (BSM) models to lie between ${\cal O} (10^{-12}) < \epsilon < {\cal O} (10^{-3})$, see Refs.~\cite{Abel:2006qt,Dienes:1996zr}. After the electroweak symmetry breaking, $Z_{\mu}$ obtains a nonzero mass $m_{Z}$ and depending on how $U(1)_d$ symmetry is broken, $Z_d^{\mu}$ will receive a nonzero mass $m_{Z_d}$ and will cause mixing with the SM neutral vector boson $Z_\mu$. The $U(1)_d$ symmetry can be broken from a spontaneous symmetry breaking term through an SM singlet scalar with a non-trivial $U(1)_d$ charge. Alternatively, $U(1)_d$ gauge boson $Z_d^{\mu}$ may acquire mass by means of the Stuckelberg mechanism~\cite{Battaglia:2003gb, Gopalakrishna:2008dv, Hewett:1988xc}. The dark $Z_d$ and SM $Z$ boson can now mix to give mass eigenstates leading to parity-violating $Z_d$ coupling given by
\begin{equation}
{\mathcal L}_{\rm int} = - \left( \epsilon\, e\, J_\mu^{em} + \epsilon_Z \frac{g}{\cos \theta_W} J_\mu^{\rm NC}\right)Z_d^\mu,
\label{lint}
\end{equation} 
where $\epsilon_Z$ depends on the $Z-Z_d$ mass mixing term~\cite{Davoudiasl:2012ag, Davoudiasl:2013aya}. Pure kinetic mixing produces an effective parity-conserving interaction (eq.~\ref{ljmu}). Mass mixing (eq.~\ref{lint}) generates parity violating couplings which are fixed. In this study we explore the Lorentz structure and the magnitude of parity violation in dark $Z_d$ coupling to SM fermions. We consider a generic structure:
\begin{equation}
{\mathcal L}_{Z_d f \bar f} = - \epsilon\, e\, \bar f \left(g_V \gamma_\mu + g_A \gamma_\mu \gamma_5 \right) f Z_d^\mu.
\label{lzdff}
\end{equation} 
In the simultaneous presence of both the vector and axial-vector couplings, parity is broken and we choose the normalization $g_V^2+g_A^2=1$. This choice is motivated by the fact that most of the experimental constraints in the literature are obtained for the case of pure kinetic mixing which corresponds to $g_V = 1$ and $g_A = 0$ in eq.~\ref{lzdff}. The dark photon ($Z_d$) interactions with the SM particles can be explored in low energy phenomenology as well as in collider experiments. Dark photons of mass up to several MeV are primarily constrained by the existing bounds from cosmology, astrophysics, and accelerator (Beam-dump) experiments (see Dark sectors 2016 Workshops Report~\cite{Alexander:2016aln,Fabbrichesi:2020wbt}). Depending on the mass of the dark photon, several channels are available for the decay. For masses up to 100~GeV, the dark photon can decay into a pair of SM fermions. Dark photon of mass of $1$~GeV has been explored in Large Hadron-electron collider (LHeC) and Future Circular Hadron electron collider (FCC-he) colliders where the signal is given by the displaced decay of long-lived photon into the charged fermions. The non-observation of the signal can exclude dark photons of mass $< 1$~GeV~\cite{DOnofrio:2019dcp}. Using polarised electron beam at HERA and at electron-ion collider (EIC) authors of Ref.~\cite{Yan:2022npz} have constrained the mixing parameter $\epsilon$ to lie between $\epsilon < 0.01 - 0.02$ for $m_{Z_d}<10$~GeV. There are also few studies on dark photon of mass $\sim 10$~GeV where LHCb provides stringent constraints on the kinetic mixing parameters $\epsilon$. %ATLAS and CMS may provide better constraints on dark photons of mass 40~GeV to 70~GeV~\cite{?}. 
For dark photon mass $20 \leq m_{Z_d} \leq 330$~GeV, the 2\,$\sigma$ exclusion limits on the mixing parameter $\epsilon \leq [10^{-3}-10^{-2}]$ can be explored at future $e^+e^-$ colliders~\cite{He:2017zzr}. The Lorentz structure of $Z_d^\mu\ell^+\ell^-$ interaction has been probed recently~\cite{Lee:2020tpn} by studying the correlation between $Z_d^\mu$ angle relative to $e^+e^-$ beam direction in the $e^+e^-$ rest frame and the $\mu^-$ angle relative to the boost direction of $Z_d^\mu$ in the $Z_d^\mu$ rest-frame based in Belle II detector. It was shown that the detection of dark $Z_d^\mu$ decay into muon-pairs in $e^+e^-$ colliders can probe the parity-violating couplings.    
CMS collaborations~\cite{Sirunyan:2019wqq} puts the most stringent constraints on the dark photon in the $\sim30-75$~GeV and $\sim110-200$~GeV mass range from searches of $Z_d$ decaying into a pair of muons.   

In the next sect.~\ref{framework}, a formalism for the analysis is discussed with definitions and observable. Simulation and observable for the study are presented in sect.~\ref{simu}. Sect.~\ref{analysis} includes the analysis details with the results of this study. Summary and discussions are followed in sect.~\ref{summary}.
%%%%%%%%%%%%%%%%%%%%%%%%%%%%%%%%%%%%%%%%%%%%%%%%%%%%%%%
\section{Framework}
\label{framework}
We base our study on LHeC environment, which employs the $E_p = 7$~TeV proton beam of the LHC and electrons from an Energy Recovery Linac (ERL) being developed for the LHeC. The choice of an ERL energy of $E_e = 60$ (120)~GeV with an available proton energy $E_p = 7$~TeV would provide a centre of mass energy of $\sqrt{s} \approx 1.3\,(1.8)$~TeV at the LHeC using the LHC protons~\cite{AbelleiraFernandez:2012cc,Bruening:2013bga,AbelleiraFernandez:2012ty}. Since $Z_d$ couples to SM particles via the interactions in Lagrangian~(\ref{lzdff}), the production of $Z_d$ in $e^- p$ collider follows through charged (neutral ({\tt NC})) currents ({\tt CC}): $e^- p \to \nu_e (e^-) Z_d\, j$. To probe the mass range of $Z_d$ as a function of $\epsilon$ in this setup, we use cross-section and asymmetries associated with polarisation observables~\cite{Boudjema:2009fz, Rahaman:2016pqj} of $Z_d$ in the decay $Z_d \to \ell^+ \ell^-$.

For a spin-$s$ particle one can construct a total of $(2 s + 1)^2 - 1 = 4 s (s + 1)$ polarisation observables. In our case, it should be eight such combinations since $s = 1$, which are related to the angular distributions of its daughter ($\ell^-$) in its rest frame. The normalised decay angular distribution of the lepton $\ell^-$ from the $Z_d$ in the rest frame of $Z_d$ is given by~\cite{Boudjema:2009fz},
\begin{align}
&\frac{1}{\sigma} \frac{d\sigma}{d\Omega_{\ell^-}^\star} = 
\frac{3}{8\pi} \left[ \left(\frac{2}{3}-(1-3\delta) \ \frac{T_{zz}}{\sqrt{6}}\right) + \alpha \ p_z
\cos\theta_{\ell^-}^\star\right.\nonumber\\
&+ \sqrt{\frac{3}{2}}(1-3\delta) \ T_{zz} \cos^2\theta_{\ell^-}^\star
\nonumber\\
&+ \left(\alpha \ p_x + 2\sqrt{\frac{2}{3}} (1-3\delta)
\ T_{xz} \cos\theta_{\ell^-}^\star\right) \sin\theta_{l^-}^\star\ \cos\phi_{\ell^-}^\star\nonumber\\
&+ \left(\alpha \ p_y + 2\sqrt{\frac{2}{3}} (1-3\delta)
\ T_{yz} \cos\theta_{\ell^-}^\star\right) \sin\theta_{\ell^-}^\star\ \sin\phi_{l^-}^\star\nonumber\\
&+ (1-3\delta) \left(\frac{T_{xx}-T_{yy}}{\sqrt{6}} \right) \sin^2\theta_{\ell^-}^\star
\cos2\phi_{\ell^-}^\star\nonumber\\
&+ \left. \sqrt{\frac{2}{3}}(1-3\delta) \ T_{xy} \ \sin^2\theta_{\ell^-}^\star\
\sin2\phi_{\ell^-}^\star \right] .
\label{eq:angular_distribution}
\end{align}
Here $\theta_{\ell^-}^\star$, $\phi_{\ell^-}^\star$ are the polar and the azimuthal orientation of $\ell^-$ from the spin-$1$ particle in the rest frame of the particle ($Z_d$) with its momentum along the $z$-direction. The quantity $\alpha$ is called analysing power, and it is related to the decay vertex structure of the particle. The quantity $\delta=0$ for massless leptonic decay.\footnote{Considering the interaction of decay vertex of a spin-1 particle $V_\mu$ to two fermions $f_1$ and $f_2$ as: $\bar{f}_1\gamma^\mu \left(C_L P_L + C_R P_R\right) f_2 V_\mu$, the two parameters $\alpha$ and $\delta$ are given as: 
\begin{align}
    \alpha = \frac{2 (C_R^2-C_L^2)\sqrt{1+(x_1^2-x_2^2)^2 - 2(x_1^2+x_2^2)}}{12C_LC_Rx_1x_2+(C_R^2+C_L^2)[2-(x_1^2-x_2^2)^2 + (x_1^2+x_2^2)]}, \notag
\end{align}
\begin{align}
    \delta = \frac{4 C_L C_R x_1 x_2 + (C_R^2+C_L^2)^2[(x_1^2+x_2^2) - (x_1^2-x_2^2)^2]}{12C_LC_Rx_1x_2+(C_R^2+C_L^2)[2-(x_1^2-x_2^2)^2 + (x_1^2+x_2^2)]}, \notag
\end{align}where $x_i = m_i/m$ ($i=1, 2$), $m$ is the mass of mother particle and $m_i$ are the mass of daughters. For details we refer to Ref.~\cite{Boudjema:2009fz}.} The vector polarisations $\vec{p}$ and independent tensor polarisations $T_{ij}$ are calculable from the asymmetries constructed from the decay angular distributions above. 

The asymmetries for the polarisations are constructed as follows~\cite{Rahaman:2016pqj}:
\begin{align}
A_i = \frac{\sigma\left( {\cal N}_i > 0 \right) - \sigma\left( {\cal N}_i < 0 \right) }{\sigma\left( {\cal N}_i > 0 \right) + \sigma\left( {\cal N}_i < 0 \right)},
\label{asym_pol}
\end{align}
where $A_i$ and the corresponding observable ${\cal N}_i$ are shown in Table~\ref{asym_obs}.
\begin{table}[t]
\centering
 \begin{tabular}{cll}
 \toprule
 &{\bf Asymmetry} ($A_i)$ & {\bf Observable} (${\cal N}_i$) \\ \toprule
1 & $A_x = \frac{3}{4}\alpha p_x$  & ${\cal N}_x = \cos\phi_{\ell^-}^\star$ \\ \hline
2 & $A_y = \frac{3}{4}\alpha p_y$ & ${\cal N}_y = \sin\phi_{\ell^-}^\star$ \\ \hline 
3 & $A_z = \frac{3}{4}\alpha p_z$ & ${\cal N}_z = \cos\theta_{\ell^-}^\star$ \\ \hline
4 & $A_{x^2-y^2} = \frac{1}{\pi}\sqrt{\frac{2}{3}}(1-3\delta) (T_{xx}-T_{yy})$ & ${\cal N}_{x^2-y^2} = \cos 2\phi_{\ell^-}^\star$ \\ \hline
5 & $A_{xy} = \frac{2}{\pi}\sqrt{\frac{2}{3}}(1-3\delta) T_{xy}$   & ${\cal N}_{xy} = \sin 2\phi_{\ell^-}^\star$ \\ \hline
6 & $A_{xz} = \frac{2}{\pi}\sqrt{\frac{2}{3}}(1-3\delta) T_{xz}$ & ${\cal N}_{xz} = \cos\theta\cos\phi_{\ell^-}^\star$ \\ \hline
7 & $A_{yz} = \frac{2}{\pi}\sqrt{\frac{2}{3}}(1-3\delta) T_{yz}$ & ${\cal N}_{yz} = \cos\theta\sin\phi_{\ell^-}^\star$ \\ \hline
8 & $A_{zz} = \frac{3}{8}\sqrt{\frac{3}{2}}(1-3\delta) T_{zz}$ & ${\cal N}_{zz} = \sin 3\theta_{\ell^-}^\star$ \\
 \toprule 
\end{tabular}
\caption{The expression for asymmetry and corresponding observable defined in eq.~(\ref{asym_pol}).} 
\label{asym_obs}
\end{table} 

In addition to these polarization asymmetries, $A_{pol}$, other asymmetries can be defined in lab-frame, $A_{Lab}$, with all possible observable. This study considered $\cos\theta_j$, $\Delta\Phi\left(\nu_e,j \right)$, $\Delta\Phi\left(\nu_e, Z_d\right)$, and $\Delta\Phi\left(j, Z_d \right)$ for {\tt CC} process. Similarly for {\tt NC} process observable with $e^-$ such as $\cos\theta_{e^-}$, $\Delta\Phi\left({e^-},j \right)$, and $\Delta\Phi\left({e^-}, Z_d\right)$ in addition with $\cos\theta_j$, $\Delta\Phi\left(j, Z_d \right)$ are considered. Here $j$ is the scattered jet, and $\Delta\Phi (m,n)$ is the azimuthal-angle difference between ($m,n$). Further, a robust $\chi^2$ analysis is performed to find the limits on $\epsilon (m_{Z_d})$ using these observable and discussed in the next section. 
\begin{figure}[t]
	\centering
	\includegraphics[width=0.35\textwidth]{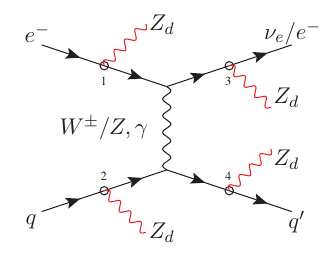}\\
	\caption{\label{feyn} Leading order representative Feynman diagrams at matrix-element level for single dark $Z_d$ production in {\tt CC} (through $W^\pm$-mediator) and {\tt NC} (through $Z$ and $\gamma$-mediators) processes through deep inelastic electron-proton collisions. The dark-$Z_d$ can radiate from either initial beams (vertex 1 or 2) or final scattered fermions (vertex 3 or 4). Here, $q, q^\prime$ $\equiv$ $u,\bar{u}$, $d,\bar{d}$, $c,\bar{c}$, $s,\bar{s}$, or $b,\bar{b}$.} 
\end{figure}
\begin{figure}[t]
	\centering
	\subfloat[\label{fig:sigma-cc}]{\includegraphics[width=0.48\textwidth]{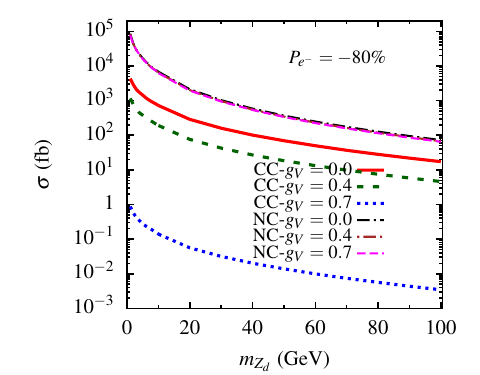}}\\
	\subfloat[\label{fig:sigma-nc}]{\includegraphics[width=0.48\textwidth]{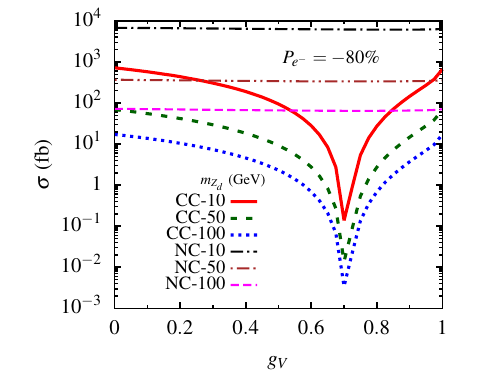}}
	\caption{\label{fig:sigma-cc-nc} The production cross-section of $Z_d$ in {\tt CC} and {\tt NC} channels including leptonic decay of $Z_d$ as a function of (a) $m_{Z_d}$ for three choices of coupling $g_V = 0.0$, $0.4$ and $0.7$, (b) $g_V$ for $m_{Z_d} = 10$, $50$ and $100$~GeV at $e^-p$ collider with $E_e = 60$~GeV and $E_p = 7$~TeV, respectively.} 
\end{figure}
%
%%%%%%%%%%%%%%%%%%%%%%%%%%%%%%%%%%%%%%%%%%%%%%%%%%%%%%%%
\begin{table}[t]
\centering
 \begin{tabular}{cc}
 \toprule
  $m_{Z_d}$ (GeV) & $m_{\ell\ell}\in[m_{\ell\ell}^{min},m_{\ell\ell}^{max}]$ (GeV) \\ \toprule
  1  & [0, 5] \\
  5  & [0, 10] \\
  10 & [0, 20] \\
  15 & [0, 30] \\
  20 & [10, 35] \\
  30 & [15, 50] \\
  40 & [20, 65] \\
  50 & [25, 80] \\
  60 & [35, 84] \\
  70 & [40, 85] \\
  80 & [50, 95] \\
  90 & [50, 100] \\
95 & [50, 190] \\
  100 & [95, 200] \\
 \toprule 
\end{tabular}
\caption{The invariant-dilepton mass cuts as discussed in the text to optimize the signal {\it vs} backgrounds for a particular mass of dark-$Z_d$ production.} 
\label{mllcut}
\end{table} 

\section{Simulation and observable}
\label{simu}
In order to explore the limits of dark-$Z_d$ we first build a model file for the Lagrangian given in eq.~(\ref{lzdff}) using the package \texttt{FeynRules}~\cite{Alloul:2013bka} and then simulate the $Z_d$ production in {\tt CC}: $p e^- \to \nu_e Z_d j$ and {\tt NC}: $p e^- \to e^- Z_d j$, with $Z_d \to \ell^+\ell^-$ pairs, in the LHeC set up as stated previously. Here we consider $\ell^\pm = e^\pm$ and $\mu^\pm$ only. The representative Feynman diagram for the production of $Z_d$ is shown in Fig.~\ref{feyn}. For the generation of events, we use the Monte Carlo event generator package \MGvATNLO~\cite{Alwall:2011uj}. The factorization and normalization scales are set to be dynamic scales for both signal and potential backgrounds. For this study, $e^-$ polarization is assumed to be $-80$\%. The initial requirements on cuts are $p_T^{\ell,j} > 10$~GeV, $|\eta_{\ell,j}| < 5$ and no cuts on missing energy.

We initiate our analysis by estimating the cross-section ($\sigma$) of $Z_d$ production as a function of its mass $m_{Z_d}$, shown in Fig.~\ref{fig:sigma-cc} for three choices of coupling, $g_V=0.0$, $0.4$, and $0.7$. The production cross-section falls as the mass of $Z_d$ increases for both {\tt CC} and {\tt NC} cases due to less available phase space energy. Variation of the cross-section due to the coupling $g_V$ is imperceptible in {\tt NC} case, while it is much prominent in the case of {\tt CC} production. The cross-section decreases from $g_V=0$ and becomes minimum around $g_V=0.7$ and increases till $g_V=1.0$ because of the constraint $g_V^2+g_A^2=1$. Further, the cross-section is smaller in {\tt CC} case than in {\tt NC} case, see Fig.~\ref{fig:sigma-nc}.

\begin{figure}[t]
\includegraphics[width=0.5\textwidth]{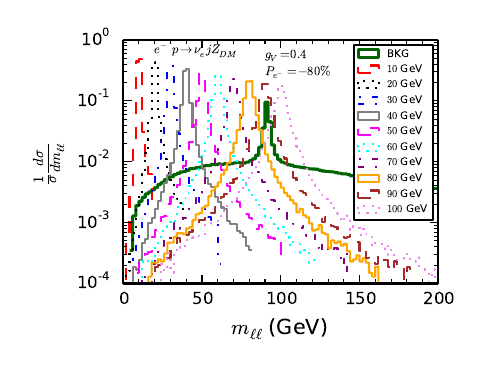}
\caption{\small The normalised distribution for the invariant mass of $\ell^+\ell^-$ pair ($m_{\ell\ell}$) are shown for the {\tt CC} process for background as well as for the signal with a representative value of $g_{V} = 0.4$. }
\label{fig:inv_mass}
\end{figure}

Next, in order to optimize the significance of the signal over all the backgrounds we chose the nominal selection cuts of $p_T^j > 20$~GeV, $p_T^\ell > 10$~GeV, $-1 < \eta_j < 5$ and $-1 < \eta_\ell < 3$. The mass dependent cut on invariant-mass distribution of di-leptonic final state (which originate from $Z_d \to \ell^+\ell^- $) $m_{\ell\ell} (m_{Z_d})$ shall be an appropriate observable to distinguish the signal from backgrounds. Note that for our signal, the background from $Z$-boson production is the dominant one, and $m_{\ell\ell} (m_{Z_d})$ cut reduces it significantly. We show the applied cuts for a  range of $Z_d$ mass in Table~\ref{mllcut}. In Fig.~\ref{fig:inv_mass}, the normalized distribution of $m_{\ell\ell}$ is shown for masses 10 to 100~GeV (in the interval of 10~GeV) with the dominant background. The peaks at $m_{Z_d}$ for signals and $m_Z$ for the background clearly show the efficacy of choosing the $m_{\ell\ell}$ cuts.  

\begin{figure}[t]
	\centering
    \subfloat[]{\includegraphics[width=0.48\textwidth]{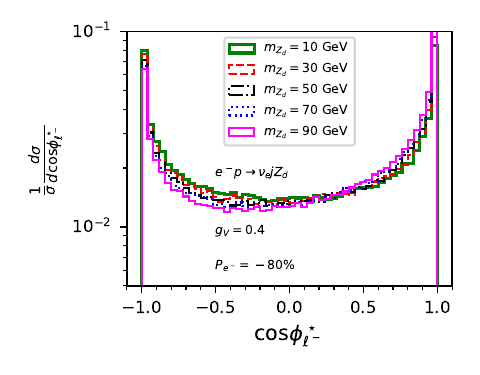}}\\
    \subfloat[]{\includegraphics[width=0.48\textwidth]{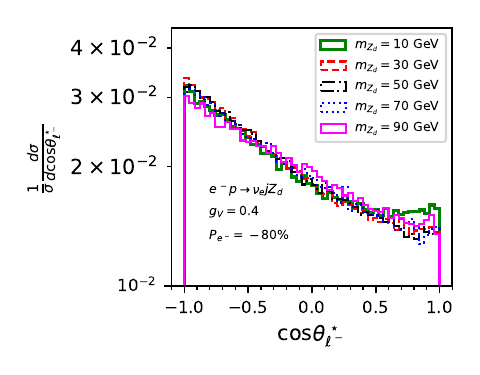}}
	\caption{\label{fig:dist-asymvarpol} The normalized distribution of $\cos\phi^\star_{\ell^-}$ and $\cos\theta^\star_{\ell^-}$ for the charged-current process with $m_{Z_d} = 10$, $30$, $50$, $70$, and $90$~GeV and $g_V=0.4$.} 
\end{figure}

\begin{figure}[t]
	\centering
    \subfloat[]{\includegraphics[width=0.48\textwidth]{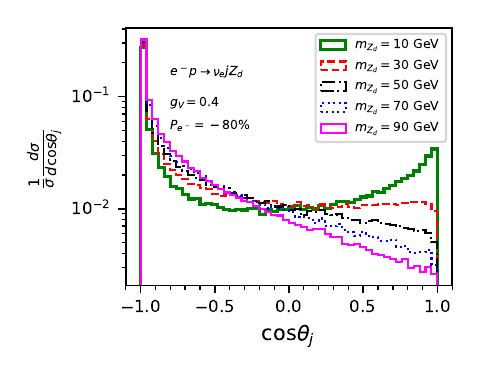}}\\
    \subfloat[]{\includegraphics[width=0.48\textwidth]{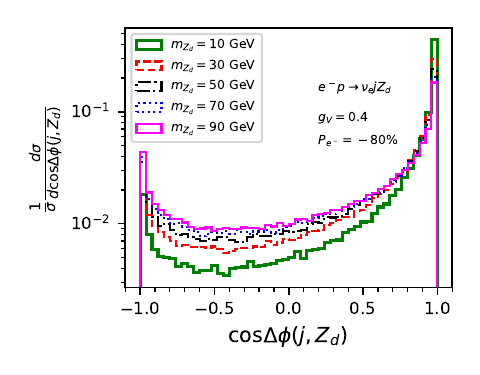}}
	\caption{\label{fig:dist-asymvar} The normalized distribution of $\cos\theta_j$ and $\cos\Delta\phi(j,Z_d)$ are shown for the charged-current process with $m_{Z_d} = 10$, $30$, $50$, $70$, and $90$~GeV and $g_V=0.4$.} 
\end{figure}

Angular observables and hence the asymmetries defined in eq.~(\ref{asym_pol}) are also significant in this study. The two angular variables ($\cos\phi_{\ell^-}^\star$, $\cos\theta_{\ell^-}^\star$) for the polarization of $Z_d$ (Fig.~\ref{fig:dist-asymvarpol}) along with two other angular variables independent of the polarization of $Z_d$ are depicted in Fig.~\ref{fig:dist-asymvar} for few benchmark of $m_{Z_d}$ and $g_V=0.4$ in the {\tt CC} process as representative distribution. The asymmetry for the $\cos\phi_{\ell^-}^\star$ becomes more positive as $m_{Z_d}$ increases, although not much change is seen for $\cos\theta_{\ell^-}^\star$. On the other hand, asymmetries for $\cos\theta_j$ and $\cos\Delta\phi(j, Z_d)$ become more negative as $m_{Z_d}$ increases. Asymmetries for these angular variables, thus, possess the potential ability to constrain on $m_{Z_d}$.


\section{Analysis and results}
\label{analysis}
After having a preliminary understanding of cross-sections of signal and backgrounds, its optimization through $m_{\ell\ell}$, and with different potential angular observables, we follow analysis using an efficient $\chi^2$-formula and describe the limits on $\epsilon$ in this section. We use observable comprising the cross-section, eight polarization asymmetries and the lab-frame asymmetries as defined in eq.~(\ref{asym_pol}) to form a total $\chi^2$ and obtain limits on the $\epsilon(m_{Z_d})$ (see eq.~(\ref{lzdff})). The total $\chi^2$ is defined as:
\begin{align}
\chi^2 = \frac{\left( \sigma(g_f) - \sigma \right)^2}{\delta\sigma^2} + \sum_j \frac{\left( A_j (g_f) - A_j \right)^2}{\delta A_j^2}.
\label{chi2}
\end{align}   
The errors $\delta\sigma$ and $\delta A_j$ in eq.~(\ref{chi2}) are obtained from the SM backgrounds and given by
\begin{align}
\delta \sigma = \sqrt{\frac{\sigma}{{\cal L}} + \left( \epsilon_\sigma \sigma\right)^2}, \quad
\delta A_j = \sqrt{\frac{1-A_j^2}{{\sigma \cal L }} + \left( \epsilon_{A_j}\right)^2},    
\end{align} 
which comprised of the integrated luminosity ${\cal L}$ dependent statistical errors in the first term and systematic errors $\epsilon_\sigma (\epsilon_{A_j})$ for the cross-section (asymmetries) in the second term. We have chosen $\epsilon_\sigma=0.02$ and $\epsilon_{A}=1~\%$ as benchmark for our analysis. 
\begin{figure}[t]
	\centering
	\subfloat[\label{cc-chi2}]{\includegraphics[width=0.48\textwidth]{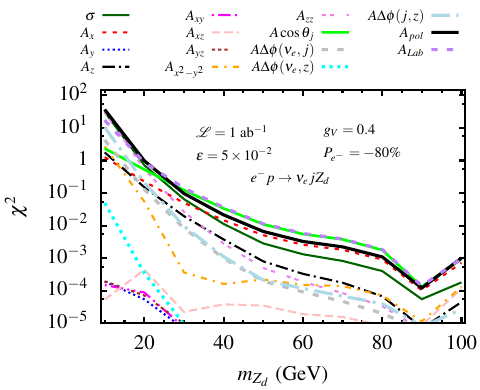}}\\
	\subfloat[\label{nc-chi2}]{\includegraphics[width=0.48\textwidth]{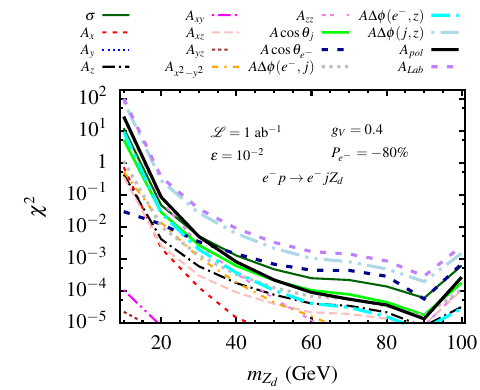}}
	\caption{\label{fig:chi2-mzd-allObs-cut2} The $\chi^2$ of cross-section ($\sigma$) and all asymmetries ($A_{pol}$ , $A_{Lab}$) are shown as a function of $m_{Z_{d}}$ for $g_V = 0.4$ with (a) $\epsilon =5\times 10^{-2}$ for the {\tt CC} process and (b) $\epsilon = 10^{-2}$ for the {\tt NC} process with integrated luminosity of ${\cal L} = 1$ ab$^{-1}$. } 
\end{figure}

\subsection{Comparing the variables based on $\chi^2$}
In order to compare the strength of various variables, we estimate the $\chi^2$ (eq.~\ref{chi2}) for the cross-section and all the angular variables separately as well as in various combinations in both {\tt CC} and {\tt NC} processes. The $\chi^2$ values for different variables are shown in Fig.~\ref{fig:chi2-mzd-allObs-cut2}  as a function of $m_{Z_d}$ with $g_V=0.4$ for an integrated luminosity of ${\cal L}=1$ ab$^{-1}$. The polarization asymmetries ($A_{pol}$) performs better than cross-section in {\tt CC} process, while it performs poorer in {\tt NC} process after $m_{Z_d}=30$~GeV. The lab frame asymmetries ($A_{Lab}$), however, perform better than both polarization asymmetries and cross-section in both processes. The $\chi^2$ values decreases as $m_{Z_d}$ increases (since the production cross-section decreases with increasing values of $m_{Z_d}$ in both processes) and it shows a dip around $m_{Z_d} \approx 90$~GeV because of $m_{\ell\ell} \approx m_Z$ invariant mass cut from the background.
\begin{figure}[t]
	\centering
	\subfloat[\label{eps-mzd-cc}]{\includegraphics[width=0.48\textwidth]{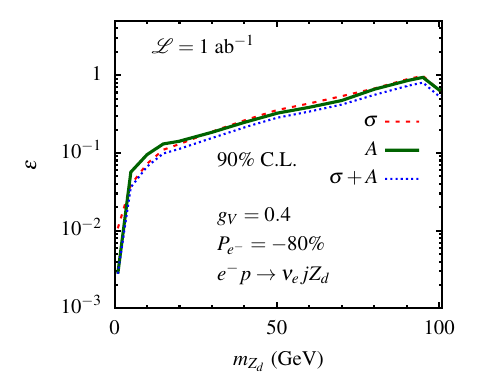}}\\
    \subfloat[\label{eps-mzd-nc}]{\includegraphics[width=0.48\textwidth]{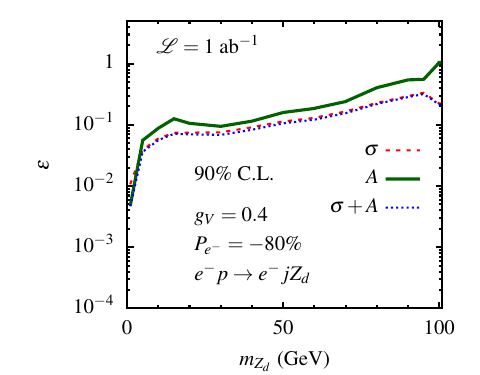}}
	\caption{\label{fig:limit-eps-mzd-cut2} The $90\%$ C.L. contours are shown in the $\epsilon$-$m_{Z_d}$ plane with the observable $\sigma$, $A$ and $\sigma + A$ for $g_V = 0.4$ for (a) {\tt{CC}} and (b) {\tt{NC}} process with integrated luminosity of ${\cal L} = 1$ ab$^{-1}$. } 
\end{figure}
\begin{figure}[t]
	\centering
	\subfloat[\label{eps-gv-cc}]{\includegraphics[width=0.48\textwidth]{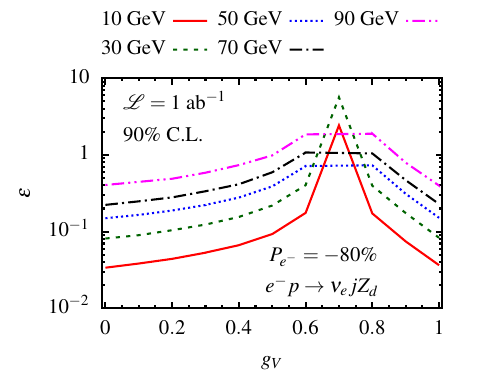}}\\
	\subfloat[\label{eps-gv-cc}]{\includegraphics[width=0.48\textwidth]{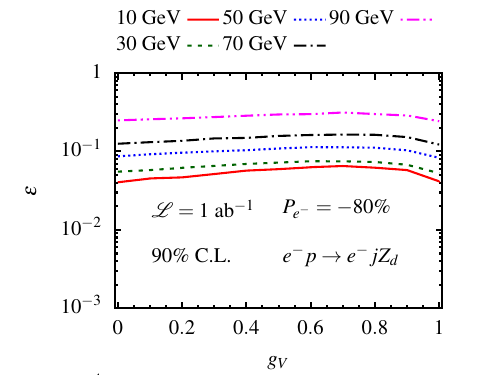}}
	\caption{\label{fig:limit-eps-gv-cut2}The $90\%$ C.L. contours are shown in the $\epsilon$-$g_V$ plane with the observable $\sigma + A$ for $m_{Z_d} = 10$, $30$, $50$, $70$, and $90$~GeV for (a) {\tt{CC}} and (b) {\tt{NC}} processes with integrated luminosity of ${\cal L} = 1$ ab$^{-1}$. } 
\end{figure}
\begin{figure}[t]
	\subfloat[]{\includegraphics[width=0.48\textwidth]{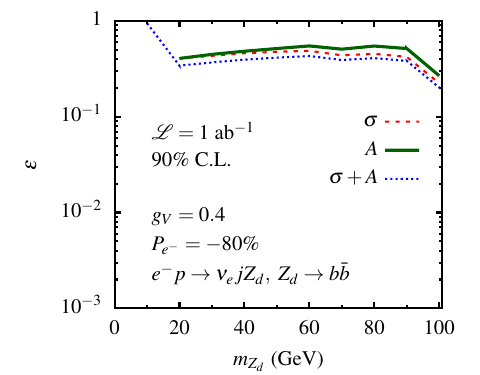}}\\
	\subfloat[]{\includegraphics[width=0.48\textwidth]{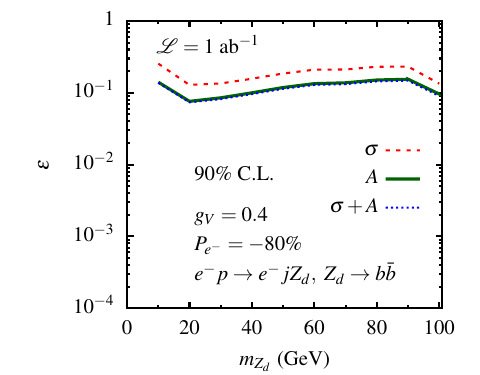}}
	\caption{\small \label{fig:limit-eps-gv-zdbb-cut2} The $90\%$ C.L. contours are shown in the $\epsilon$-$m_{Z_d}$ plane with the observable $\sigma$, $A$ and $\sigma + A$ for $g_V = 0.4$ with $Z_d\to b\bar{b}$ for both (a) charged current and (b) neutral current process with integrated luminosity of ${\cal L} = 1$ ab$^{-1}$. } 
\end{figure}
\begin{figure*}[t]
\begin{center}
\includegraphics[width=0.75\textwidth]{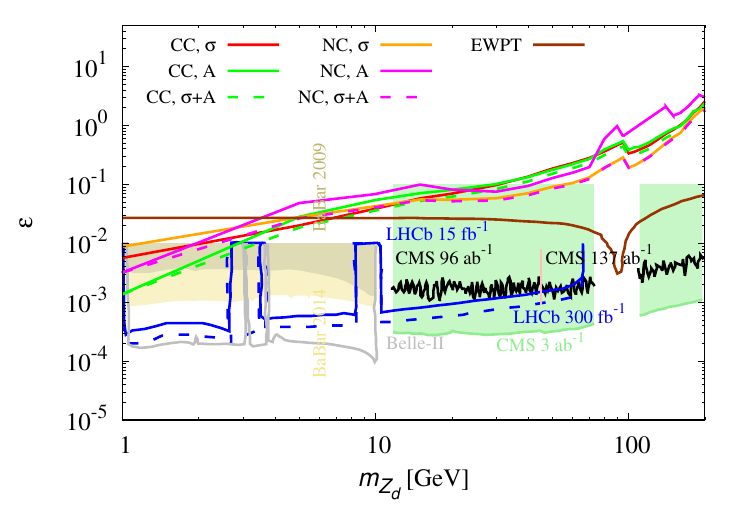}
\caption{\small The $90\%$ C.L. contours  in the $\epsilon$-$m_{Z_d}$ plane are shown 
	with the observable $\sigma$, $A$ and $\sigma + A$ for $g_V = 1.0$ for both
	charged current process as well as neutral current process with an integrated luminosity of ${\cal L} = 1$ ab$^{-1}$. Corresponding available limits from various experiments such as LHCb, BaBar, Bell-II and CMS are also shown for comparison. The light-green solid line shows the projection of the CMS experiment results at $\sqrt{s} = 13$~TeV from dimuon channel for $\mathcal{L} = 3$~ab$^{-1}$~\cite{Hosseini:2022urq}. Constraints at 95\% C.L. from the measurements of the electroweak observables (EWPT) are shown in brown color. To illustrate the trend beyond $m_{Z_d} = 100$~GeV for the same parameters, we have also shown the limits up to $m_{Z_d} \le 200$~GeV.}
\label{fig:constraints_exp}
\end{center}
\end{figure*}
\begin{figure}[t]
\begin{center}
\includegraphics[width=0.48\textwidth]{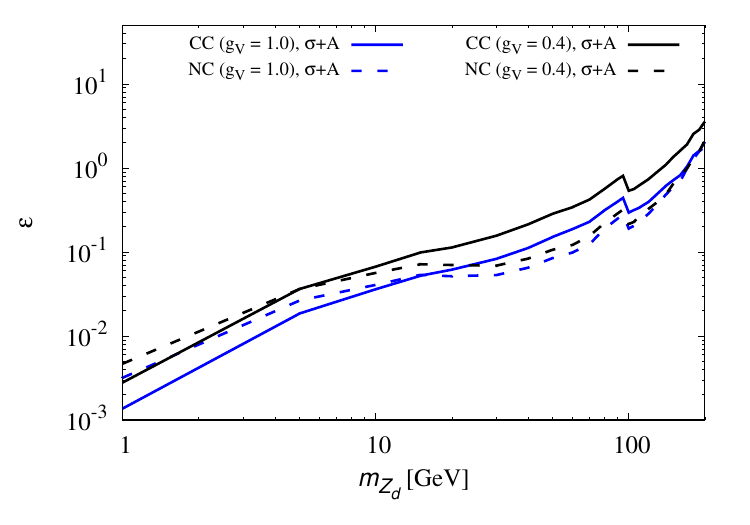}
\caption{\small A comparison plot on limits of $\epsilon-m_{Z_d}$ plane for $g_V = 0.4$ with the $g_V = 1.0$ case at 90\% C.L.}
\label{fig:compare}
\end{center}
\end{figure}

\subsection{Limits on $\epsilon$ as function of $m_{Z_d}$ and $g_V$}

We obtain two parameter limits on the $\epsilon$-$m_{Z_d}$ and $\epsilon$-$g_V$ plane by fixing the values of $g_V$ and $m_{Z_d}$ respectively. In case of $\epsilon$-$m_{Z_d}$ plane we consider the observable as (a) cross-section, (b) all asymmetries together ($A \equiv \sum_i A_i + A_{pol}+A_{Lab}$), and (c) cross-section along with asymmetries in both {\tt CC} and {\tt NC} processes. The $90\%$ confidence level (C.L.) contours ($\chi^2=4.61$~\cite{Cowan:2010js}) for two parameter in the $\epsilon$-$m_{Z_d}$ plane are shown in Fig.~\ref{fig:limit-eps-mzd-cut2} with $g_V=0.4$ for both {\tt CC} (Fig.~\ref{eps-mzd-cc}) and {\tt NC} (Fig.~\ref{eps-mzd-nc}) processes with an integrated luminosity of ${\cal L}=1$ ab$^{-1}$. We scanned the mass of $Z_d$ in the range of $0.1$ - $100$~GeV for cross-section. For the asymmetries, however, we scanned $m_{Z_d}$ in the range of $1 - 100$~GeV (due to the limitation in event generation). The limits on $\epsilon$ get weaker with increasing $m_{Z_d}$ up to $90$~GeV as expected from the Fig.~\ref{fig:chi2-mzd-allObs-cut2}. The asymmetry observables provide better limits on $\epsilon$ compared to the cross-section for the studied mass range. Also the limits in {\tt NC} process are better compared to {\tt CC} process because of larger cross-sections (see Fig.~\ref{fig:sigma-cc-nc}).

Similarly in Fig.~\ref{fig:limit-eps-gv-cut2}, we show the limits on $\epsilon$ as a function of $g_V$ for fixed $m_{Z_d} = 10$, $30$, $50$, $70$ and $90$~GeV by considering cross-section along with all asymmetries in both {\tt CC} and {\tt NC} processes. Since the cross-section for {\tt CC} process is minimal around $g_V \approx 0.7$ (Fig.~\ref{fig:sigma-cc}), limits on $\epsilon$ get weaker as it approaches to $g_V \approx 0.7$ and get stronger afterwards. Due to the cross-section behaviour seen in Fig.~\ref{fig:sigma-nc} for {\tt NC} process there is almost no fluctuation on $\epsilon$. Since the value of total cross-section decreases as a function of increasing $m_{Z_d}$, the limits on $\epsilon$ also get weaker as $m_{Z_d}$ goes higher in both {\tt CC} and {\tt NC} processes for full range of $g_V$.

%%%%%%%%%%%%%%%%%%%%%%%%%%%%%%%%%%%%%%%%
\subsection{Analysis in $Z_d\to b\bar{b}$ channel}

At this point, it is prudent to ask if the limits obtained in $Z_d \to \ell^+\ell^-$ decay modes are comparable with $Z_d \to jj$. However, in this decay mode, the polarization observable can not be constructed. So we only consider cross-section and the lab frame asymmetries for  the $Z_d \to b \bar{b}$ decay in both {\tt CC} and {\tt NC} processes. And we repeat the analysis to extract limits on the new physics parameter ($m_{Z_d}$, $\epsilon$, $g_V$) using the $\chi^2$-method (eq.~\ref{chi2}) in this channel. 

The corresponding $90\%$ C.L. contours in the $\epsilon$-$m_{Z_d}$ plane are shown in Fig.~\ref{fig:limit-eps-gv-zdbb-cut2} with $g_V=0.4$ for the same integrated luminosity of ${\cal L}=1$~ab$^{-1}$ for both {\tt CC} and {\tt NC} processes. The order of $\epsilon (m_{Z_d})$ remains roughly the same with a little variation of similar nature as in Fig.~\ref{fig:limit-eps-mzd-cut2}. But the limits are much weaker in this channel because of the large $Z\to b\bar{b}$ background compared to the $Z_d\to \ell^-\ell^+$ channel.
%
%%%%%%%%%%%%%%%%%%%%%%%%%%%%%%%%%%%%%%%%%%%%%%%%%%%%%%%
\section{Summary and discussions}
\label{summary}
In this article, we consider a vector-portal interaction in which the dark and visible sectors interact through kinetic mixing, leading to a so-called heavy dark photon, $Z_d$. To explore the Lorentz structure and strength of $Z_d$ coupling to fermions we assume a generic structure involving two set of couplings $\epsilon,\, g_V$ (where $g_V^2 + g_A^2 = 1$) and explore the limits for $m_{Z_d} \in [\sim 1-100]$~GeV in LHeC environment. A particular decay mode $Z_d \to \ell^+\ell^-$ has been taken for the study in which an invariant mass $m_{\ell\ell}$ cut significantly able to reduce the dominant background from $Z$-boson. In order to obtain limits on $\epsilon$-$m_{Z_d}$ and $\epsilon$-$g_V$ planes, a $\chi^2$-method is used in which different observable $viz.$ cross-section, polarisation asymmetries, and Lab-frame asymmetries are taken as input. Interestingly the charged-current process shows large variations in cross-section with $g_V$ in comparison to neutral-current process at the LHeC $\sqrt{s} \approx 1.3$~TeV considered in this study. Though the higher cross-section of {\tt NC} channel leads to stronger limits on $\epsilon$ in comparison to {\tt CC} channel. 
The main results of our study to explore the Lorentz structure of the interaction between dark photon and fermions are shown in Fig.~\ref{fig:limit-eps-mzd-cut2} and~\ref{fig:limit-eps-gv-cut2} where 90\% C.L. contours are drawn in the $\epsilon$-$m_{Z_d}$ and $\epsilon$-$g_V$ plane for representative sets of coupling $g_V$ and $m_{Z_d}$, respectively.

A summary plot on limits of $\epsilon$-$m_{Z_d}$ with $g_V = 1.0$ at 90\% C.L. is shown in Fig.~\ref{fig:constraints_exp}, where the limits are compared and depicted with existing experiments (for available dark photon mass) from CMS~\cite{Sirunyan:2019wqq}, LHCb~\cite{Aaij:2019bvg}, Belle-II~\cite{Kou:2018nap} and BaBar~\cite{Lees:2014xha}. Constraints at 95\% C.L. from the measurements of the electroweak observables are also shown in brown color~\cite{Curtin:2014cca}. For a low mass of order $m_{Z_d}< 10$~GeV, the limits of $\epsilon$ are nearly within the limits of BaBar and LHCb experiments. But for the higher masses $m_{Z_d}>10$~GeV the LHeC with $\sqrt{s}\approx 1.3$~TeV results weaker limits of $\epsilon$-$m_{Z_d}$ in comparison to these experiments. In Fig.~\ref{fig:compare}, we compare the limits of $\epsilon$-$m_{Z_d}$ for $g_V = 0.4$ with the $g_V = 1.0$ case at 90\% C.L.

To summarise our study, we also compared the limits on $\epsilon$-$m_{Z_d}$ plane for $Z_d \to b \bar{b}$ which gives a little weaker limits in comparison to $Z_d \to \ell^+\ell^-$ due to the large background. Interestingly it is noted that in the $\epsilon$-$m_{Z_d}$ plane, the limits get weaker as $m_{Z_d}$ approaches the $Z$-boson mass and get stronger after crossing this mass in all cases. And thus, the SM background due to $Z$-boson plays a very significant role within $m_{Z_d} \approx 100$~GeV. 

To our knowledge, this is the first study we carried out in the LHeC environment to obtain limits on the searches of high mass ($> 1$~GeV) dark photon with sensitive observable, where without evading the theoretical constraint on the limit of $\epsilon\lesssim\mathcal{O}(1)$, $m_{Z_d}$ up to $100$~GeV can be explored with $\sqrt{s} \approx 1.3$~TeV energy. By increasing the energy of electron sufficiently in the LHeC, higher masses of $m_{Z_d} > 100$~GeV can be studied, but then the single Higgs-boson (of mass $125$~GeV) production in both charged- and neutral-current will be the dominant SM backgrounds in $Z_d\to b\bar{b}$ decay, and hence appropriate $m_{\ell\ell}$ or $m_{jj}$ cut will help to reduce this background and limits on $\epsilon$-$m_{Z_d}/g_V$ can be obtained accordingly.

%%%%%%%%%%%%%%%%%%%%%%%%%%%%%%%%%%%%%%%%%%%%%%%%%%%%%%%
\section*{Acknowledgment}
AG thanks SERB, G.O.I. under CRG/2018/004889.
%We acknowledge fruitful discussions ....

\end{document}